\documentclass[pss,fleqn,embeddedheads]{w-art}
\usepackage{times}
\usepackage{w-thm}
\usepackage[]{graphicx}

\begin{document}
%%    The information for the title page will be placed between
%%    \begin{document} and \maketitle. The order of most entries
%%    is determined by the class file and can not be changed by
%%    rearranging them. The maketitle command follows after the
%%    abstract.
%%
%%    Most of the following commands will be completed by the publisher.
%%
%%    The copyrightyear is defined in the .clo file as the first argument
%%    of the copyrightinfo command. If the copyrightyear differs from that
%%    value it might be adjusted by the following definition:
%%
%% \renewcommand{\copyrightyear}{2003}% uncomment to change the copyrightyear.
%%

\renewcommand{\vec}[1]{\ensuremath{{\bf #1}}}

\DOIsuffix{theDOIsuffix}
%%
%% issueinfo for header and copyright line
\Volume{XX}
\Issue{1}
\Month{01}
\Year{2003}
%%
%%    First and last pagenumber of the article. If the option
%%    'autolastpage' is set (default) the second argument may be left empty.
\pagespan{3}{}
%%
%%    Dates will be filled in by the publisher. The 'reviseddate' and
%%    'dateposted' (Published online) entry may be left empty.
\Receiveddate{21 August 2005}
\Reviseddate{27 August 2005}
\Accepteddate{}
\Dateposted{$Revision: 1.13 $, compiled \today}

%% \pretitle{Editor's Choice}

%% We have a short and a long form for the title. The short form
%% (optional argument) goes into the running head.

%\title{Electron-Electron Interactions and the Integer Quantum Hall Effect}
\title[Compressibility in IQHE within HF approximation]%
{Compressibility in the Integer Quantum Hall Effect within Hartree-Fock Approximation}

%% Please do not enter footnotes or \inst{}-notes into the optional
%% argument of the author command. The optional argument will go into
%% the header.  If there is only one address the marker \inst{x} may be
%% omitted.

%% Information for the first author.
\author[Christoph Sohrmann]{Christoph Sohrmann\footnote{Corresponding
     author: e-mail: {\sf c.sohrmann@warwick.ac.uk}, Phone: +44\,2476\,574\,309,
     Fax: +44\,7876\,858\,246}\inst{1}}
%%
%%    Information for the second author
\author[Rudolf A.\ R\"{o}mer]{Rudolf A.\ R\"{o}mer\inst{1}}
\address[\inst{1}]{Physics Department \& Centre for Scientific Computing,
                   University of Warwick, Coventry CV4 7AL, U.K.}
%%
%%    Information for the third author
%%    \author[...]{... \inst{2}}
%%
%%    \dedicatory{This is a dedicatory.}

%% ABSTRACT
\begin{abstract}
  Electron-electron interactions seem to play a surprisingly small role
  in the description of the integer quantum Hall effect, considering
  that for just slightly different filling factors the interactions are
  of utmost importance causing the interaction-mediated fractional
  quantum Hall effect.  However, recent imaging experiments by Cobden et
  al.\ \cite{CobBF99} and Ilani et al.\ \cite{IlaMTS04} constitute
  strong evidence for the importance of electron-electron interactions
  even in the integer effect.  The experiments report on measurements of
  the conductance and electronic compressibility of mesoscopic MOSFET
  devices that show disagreement with predictions from the single
  particle model.  By diagonalising a random distribution of Gaussian
  scatterers and treating the interactions in Hartree-Fock approximation
  we investigate the role of electron-electron interactions for the
  integer quantum Hall effect and find good agreement with the experimental
  results.
\end{abstract}
%% maketitle must follow the abstract.
\maketitle                   % Produces the title.

%% If there is not enough space inside the running head
%% for all authors including the title you may provide
%% the leftmark in one of the following three forms:

%% \renewcommand{\leftmark}
%% {First Author: A Short Title}

%% \renewcommand{\leftmark}
%% {First Author and Second Author: A Short Title}

%% \renewcommand{\leftmark}
%% {First Author et al.: A Short Title}

%% \tableofcontents  % Produces the table of contents.

%%%%%%%%%%%%%%%%%%%%%%%%%%%%%%%%%%%%%%%%%%%%%%%%%%%%%%%%%%%%%%%%%%%%%%%%%%%%%%%
%%
%% INTRODUCTION
%%
%%%%%%%%%%%%%%%%%%%%%%%%%%%%%%%%%%%%%%%%%%%%%%%%%%%%%%%%%%%%%%%%%%%%%%%%%%%%%%%
\section{Introduction}
\label{SecIntroduction}

The integer quantum Hall effect (IQHE) --- observed in two-dimensional
electron systems (2DES) subject to a strong perpendicular magnetic field
$B$ \cite{KliDP80} --- has been explained in great detail based on
single-particle arguments \cite{Pru87}. There exists one extended
state in the centre of each disorder-broadened Landau level where the
accompanying localisation-delocalisation transition is governed by the
critical exponent $\tilde{\nu}=2.34 \pm 0.04$ \cite{HucK90,KocHKP91b}, the
Hall conductivity $\sigma_{xy}$ jumps by $e^2/h$ and the longitudinal
conductivity $\sigma_{xx}$ is finite.

However, recent experiments on mesoscopic MOSFET devices questioned that
simple picture. Measurements of the Hall conductance as a function of
magnetic field $B$ and gate voltage \cite{CobBF99} exhibited regular
patterns along integer filling factors. It was argued that, contrary to
the single-particle picture, these patterns should be attributed to Coulomb
blockade effects.
Similar patterns have been found recently also in measurements
of the electronic compressibility $\kappa$ as a function of $B$ and
electron density $n$ \cite{IlaMTS04}. From these measurements it turns out that the
number of localised states is independent of $B$ which is inconsistent with the single-particle picture.
The authors explain these results by the incomplete screening of the impurity charge density at the Landau
level band edges.

In the present paper, we investigate the effects of Coulomb
interactions on compressibility and Hall conductance within a
Hartree-Fock (HF) approach. We show that the observed charging lines in
the compressibility can be found within HF. Similar approaches
\cite{StrK05,PerC05} have recently found supporting results.

%%%%%%%%%%%%%%%%%%%%%%%%%%%%%%%%%%%%%%%%%%%%%%%%%%%%%%%%%%%%%%%%%%%%%%%%%%%%%%%
\section{Hartree-Fock approximation in the Landau basis}

We consider a 2DES confined to a square of size $L \times L$ with
periodic boundary conditions. The Coulomb interaction is treated within
the Hartree-Fock approximation \cite{Aok79,MacG88,MacA86} and the resulting Hartree-Fock
Hamiltonian is given as
\begin{align}
  H_{\rm HF}^{\sigma} = H_0^\sigma + V_{\rm HF}^\sigma = \frac{(\vec{p}-e\vec{A})^2}{2m^*} + \frac{\sigma g^* \mu_{\rm B} B}{2}
  + V_{\rm D}(\vec{r}) + V_{\rm HF}^{\sigma}(\vec{r}),
\end{align}
where $\sigma=\pm 1$ denotes the spin and $V_{\rm D}$ is a smooth
disorder potential.
The kinetic term in the Hamiltonian can be diagonalised by the Landau
wave functions
\begin{align}
  \langle\vec{r}|\phi_{m,k}\rangle = \sum_{u=-\infty}^{\infty}
  \frac{1}{\sqrt{2^m m! \pi^{\frac{1}{2}}l_{\rm c}L}} \exp\left[{\rm i} k
  y-\frac{(x-k l_{\rm c}^2+u L)^2}{2l_{\rm c}^2}\right] H_m\left(\frac{x-kl_{\rm c}^2+u L}{l_{\rm c}}\right),
\end{align}
where $l_{\rm c}=\sqrt{\hbar/eB}$ is the magnetic length and the Landau
gauge $\vec{A} = Bx\vec{e}_y$ has been used. The Landau level index is denoted by $m$
and the momentum $k= 0, 1, \ldots, N_{\phi}-1$, where the number of
flux quanta is related to $L$ and $l_{\rm c}$ by $N_{\phi}= {L^2}/{2 \pi l_{\rm c}^2}$.
The potential $V_{\rm D}$ is constructed as a sum of
$N_s$ Gaussian scatterers at random positions $\vec{r}_s$ with random
strength $V_s$ and fixed range $d$ within the square,
\begin{align}
  V_{\rm D}(\vec{r}) = \sum_{u=-\infty}^{\infty}\sum_{s=1}^{N_s}
  \frac{V_s}{\sqrt{2\pi d^2}}\exp\left[\frac{(\vec{r}-\vec{r}_s+u L \vec{e}_x)^2}{2d^2}\right] .
\end{align}
The electron-electron interactions are treated in Hartree-Fock
approximation resulting in an effective Coulomb and exchange potential
$V_{\rm HF}^{\sigma}$ which has to be calculated self-consistently.  On
solving the (self-consistent) eigenvalue equation ${\bf H}_{\rm HF}^{\sigma}
{\bf C}_\alpha^{\sigma} = \epsilon_\alpha^{\sigma} {\bf C}_\alpha^{\sigma}$, we
obtain the single particle energies $\epsilon_\alpha^{\sigma}$ and the
expansion coefficients $C_{\alpha i}^\sigma$ of the HF wave functions
$|\psi_{\alpha}\rangle = \sum_{\sigma,i} C_{\alpha i}^\sigma
|\phi_i\rangle$. \\
The total energy of the system is then calculated as
\begin{align}
  E_{N(\epsilon_{\rm F})}^{\rm tot} &=  \frac{1}{2}\sum_{\sigma,i,j}\varrho^\sigma_{ij}\langle\phi_i|2 H^\sigma_0 + V^{\sigma}_{HF}|\phi_j\rangle \quad
\end{align}
with the density matrix
%  \mbox{with the density matrix}
\begin{align}
  \quad \varrho^{\sigma}_{a b} = \sum^{\epsilon^{\sigma}_{\alpha}\leq \epsilon_{\rm F}}_{\alpha} \left(C^{\sigma}_{\alpha a}\right)^{*}C_{\alpha b}^\sigma \quad .
\end{align}
Here $i$, $j$ denote multi-indices counting both Landau level index $m$
and momentum $k$.

%%%%%%%%%%%%%%%%%%%%%%%%%%%%%%%%%%%%%%%%%%%%%%%%%%%%%%%%%%%%%%%%%%%%%%%%%%%%%%%
\section{Electronic Compressibility and Hall Conductivity}

%\section{Electronic Compressilibity}

\begin{figure}[!ht]
    \center
    \begin{minipage}[t]{.48\textwidth}
        \includegraphics[width=\textwidth]{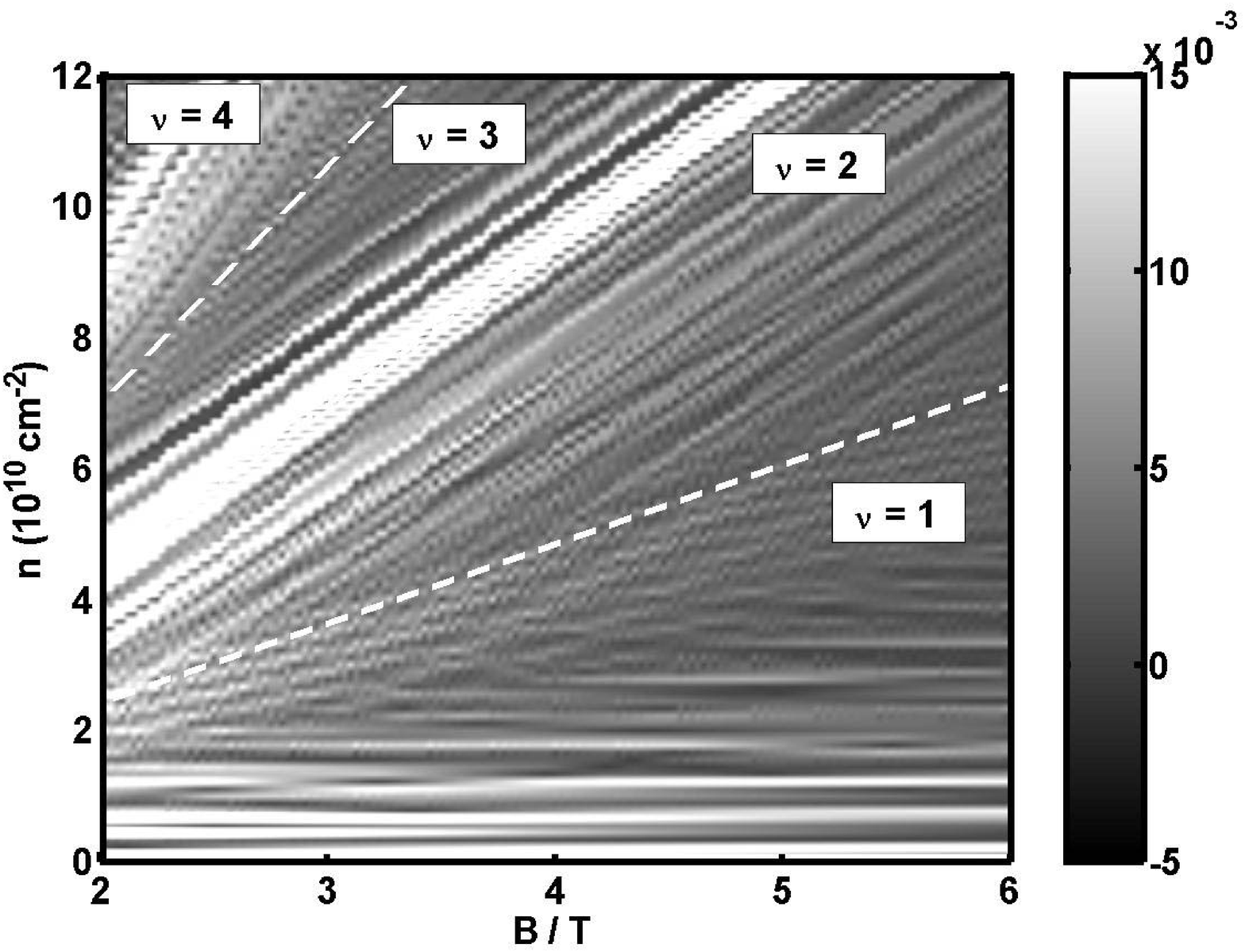}
        \caption{
        Inverse compressibility of a non-interacting system as a function of $B$ and $n$
        for the lowest $2$ Landau levels.}
        \label{fig:compress-no-interaction}
    \end{minipage}\hfil
    \begin{minipage}[t]{.48\textwidth}
        \includegraphics[width=\textwidth]{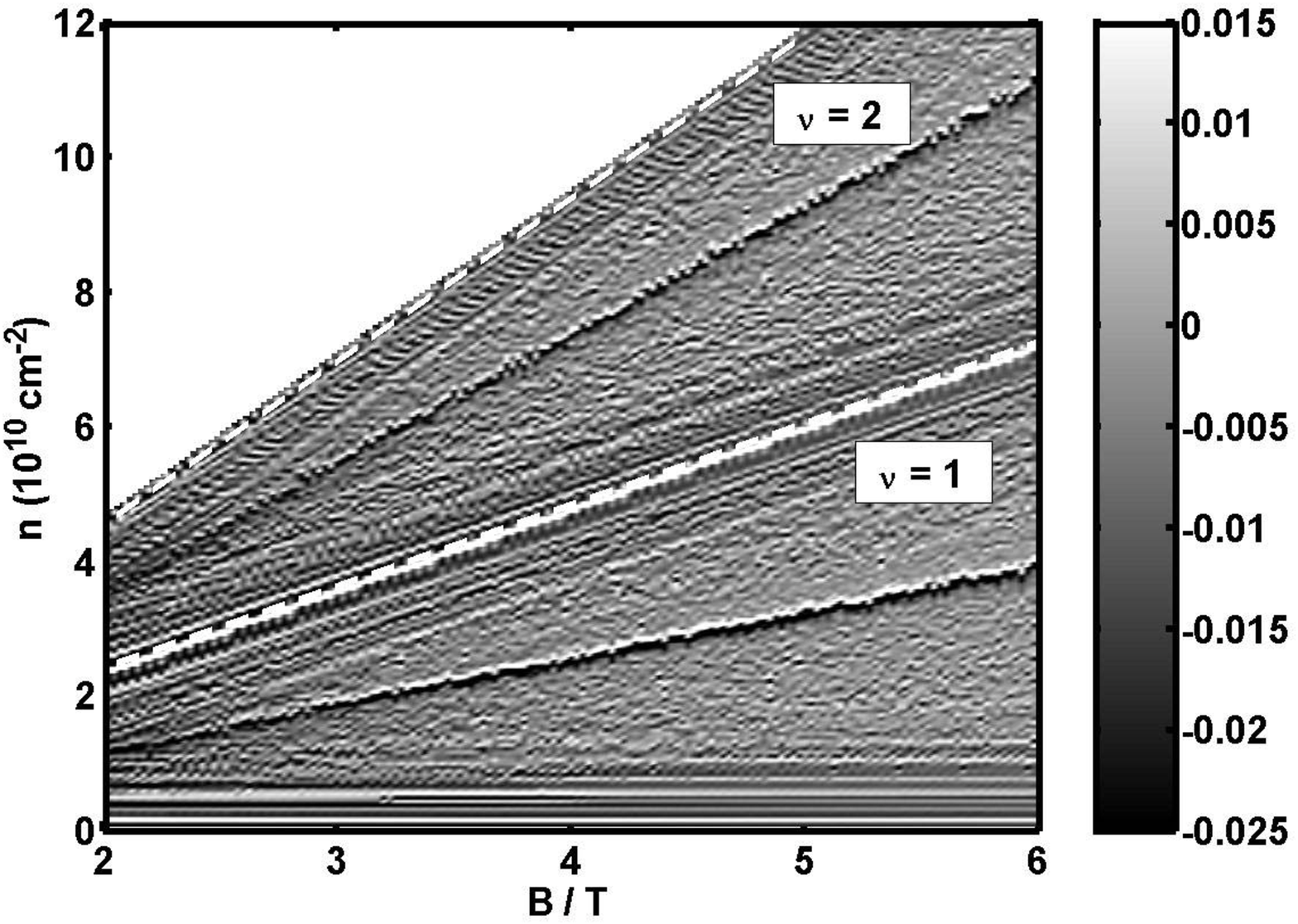}
        \caption{
        Inverse compressibility in the lowest spin-split Landau level of the same system as in
        Fig.\ \ref{fig:compress-no-interaction} but with full Coulomb interaction.}
        \label{fig:compress-yes-interaction}
    \end{minipage}
\end{figure}

The electronic compressibility $\kappa$ is inversely proportional to the linear
response of the chemical potential $\mu$ to changes in the electron density $n$,
\begin{align}
\kappa \propto \left(\frac{d\mu}{dn}\right)^{-1} \quad \mbox{with} \quad
\frac{d\mu}{dn} = L^2 \left(E_{N+1}-2 E_{N}+E_{N-1}\right),
\end{align}
where $E_N$ is the total energy of $N=n L^2$ electrons. Calculations of the
electronic compressibility are shown in Figs.\ \ref{fig:compress-no-interaction}
and \ref{fig:compress-yes-interaction} for non-interacting and interacting
electron system in the same disorder configuration, respectively. The sample size
in  both cases is $300$ nm which corresponds to $N_{\phi}=44$ at $B=2$ T and
$N_{\phi}=131$ at $B=6$T.
For the calculations we have used $N_s=200$ scatterers with range $d=20\mbox{
nm}$ and strengths $V_s/$nm randomly distributed in $[-20, 20]$ meV. Only the lowest
Landau level has been included into the calculation for the interacting system
due to the large computational effort involved.
%Spin has been neglected for clarity.

\begin{figure}[!ht]
    \center
    \begin{minipage}[t]{.48\textwidth}
        \includegraphics[width=\textwidth]{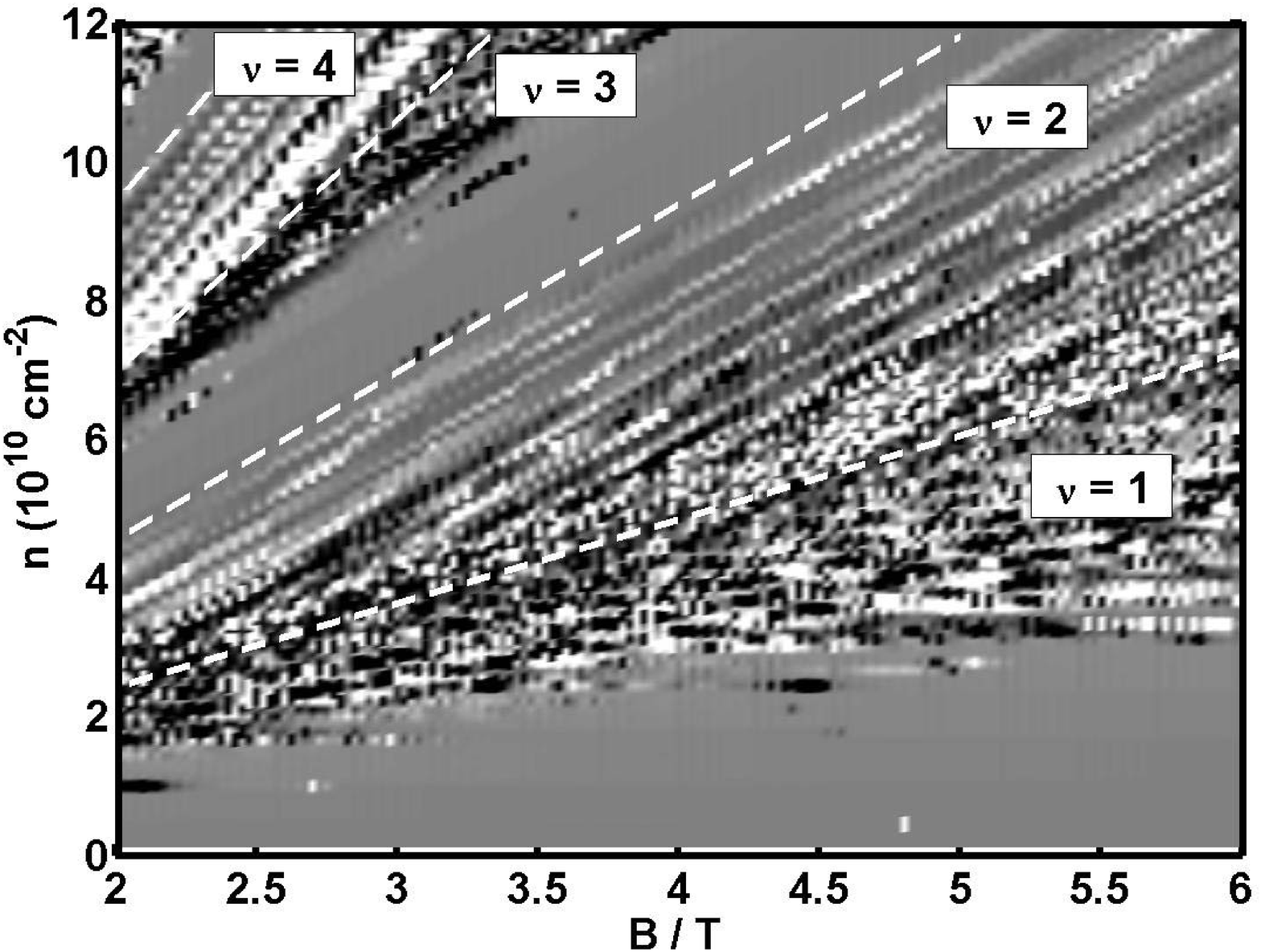}
        \caption{
        Hall conductivity $\sigma_{xy}$ as a function of $B$ and $n$ for the lowest
        $2$ Landau levels and the same disorder configuration as in Figs.
        \ref{fig:compress-no-interaction} and \ref{fig:compress-yes-interaction}.
        For clarity, we have subtracted the values of the plateau conductivities.}
        \label{fig:Conductivity2D}
    \end{minipage}\hfil
    \begin{minipage}[t]{.48\textwidth}
        \includegraphics[width=\textwidth]{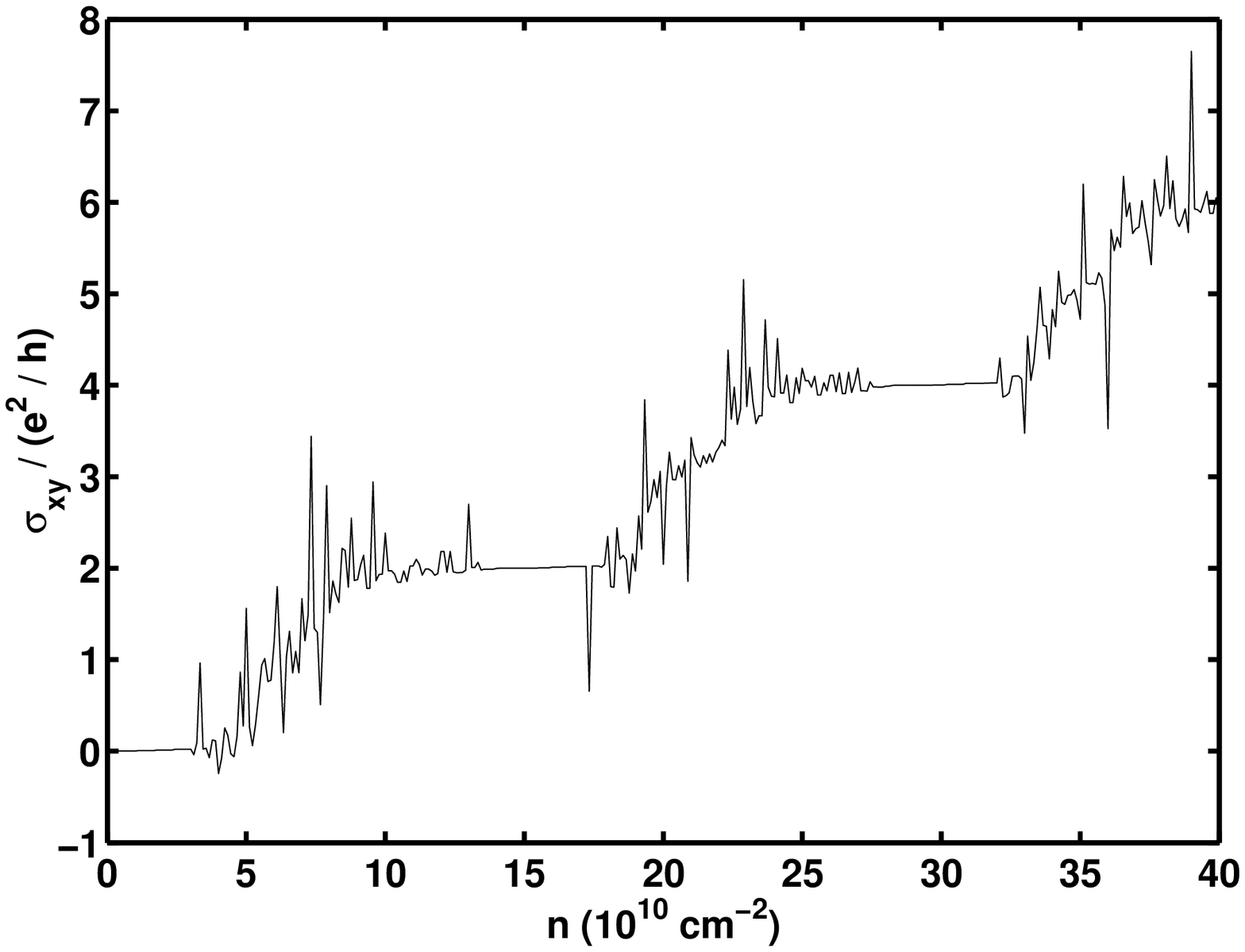}
        \caption{
        $\sigma_{xy}$ as a function of electron density $n$ at $B=5.9$T for the
        lowest $3$ Landau levels. Same sample as in Fig.\ \ref{fig:compress-no-interaction} and \ref{fig:compress-yes-interaction}.}
        \label{fig:Conductivity1D}
    \end{minipage}
\end{figure}

Without interactions the Hall conductivity can be calculated using the Kubo
formula \cite{KubMH65}. In the strong field limit, the electron motion can be
separated into guiding centre and cyclotron motion and the Kubo formula reads
\begin{align}
\sigma_{xy} &= -\frac{n ec}{B}+\frac{e^2 \hbar}{{\rm i} L^2}
\sum^{(<\epsilon_{\rm F})}_{\alpha} \sum^{(>\epsilon_{\rm F})}_{\beta}
               \frac{
                     \langle\psi_\alpha|\dot{X}|\psi_\beta\rangle
                     \langle\psi_\beta|\dot{Y}|\psi_\alpha\rangle
                     -
                     \langle\psi_\alpha|\dot{Y}|\psi_\beta\rangle
                     \langle\psi_\beta|\dot{X}|\psi_\alpha\rangle
                    }{(\epsilon_\alpha-\epsilon_\beta)^2}
                    \label{eq-simgaxy}
\end{align}
with the velocity of the guiding centres given as
\begin{align}
 \dot{X} = \frac{l_c^2}{\hbar}\frac{\partial V_{\rm D}}{\partial y}, \qquad \dot{Y} = - \frac{l_c^2}{\hbar}\frac{\partial V_{\rm D}}{\partial x}.
\end{align}
In the Landau basis, the velocity matrix elements can be written as
\begin{align}
 \langle\phi_{n,i}|\dot{X}|\phi_{m,j}\rangle
   = \frac{{\rm i} l_c^2}{\hbar} (k_i-k_j)\langle\phi_{n,i}|V_{\rm D}|\phi_{m,j}\rangle
   \label{eq-pXp}
\end{align}
and
\begin{align}
 \langle\phi_{n,i}|\dot{Y}|\phi_{m,j}\rangle
   = \frac{l_c^2}{\hbar}
    &\left(
     \sqrt{\frac{n}{2}} \langle\phi_{n-1,i}|V_{\rm D}|\phi_{m,j}\rangle
    +\sqrt{\frac{m}{2}} \langle\phi_{n,i}|V_{\rm D}|\phi_{m-1,j}\rangle
    \right. \nonumber \\
    &\left.
    -\sqrt{\frac{n+1}{2}} \langle\phi_{n+1,i}|V_{\rm D}|\phi_{m,j}\rangle
    -\sqrt{\frac{m+1}{2}} \langle\phi_{n,i}|V_{\rm D}|\phi_{m+1,j}\rangle
     \right).
     \label{eq-pYp}
\end{align}
In order to calculate $\sigma_{xy}$ for the interacting HF system, we interpret the
$V_{\rm HF}^{\sigma}$ as an additional, mean-field potential \cite{MacG88,YanMH95} and
compute $\sigma_{xy}$ using Eqs.\ (\ref{eq-pXp}) and (\ref{eq-pYp}) with
$V_{\rm D}\rightarrow V_{\rm D}+V_{\rm HF}$.
Results on the Hall conductivity in the non-interacting case are depicted in
Figs.\ \ref{fig:Conductivity2D} and \ref{fig:Conductivity1D} using the same
parameters as in the compressibility calculations.

%%%%%%%%%%%%%%%%%%%%%%%%%%%%%%%%%%%%%%%%%%%%%%%%%%%%%%%%%%%%%%%%%%%%%%%%%%%%%%%
%%
%% CONCLUSIONS
%%
%%%%%%%%%%%%%%%%%%%%%%%%%%%%%%%%%%%%%%%%%%%%%%%%%%%%%%%%%%%%%%%%%%%%%%%%%%%%%%%
\section{Discussions and Conclusions}\label{SecConclusions}

From our compressibility results for the non-interacting system, we find strongly
incompressible lines along {\em even} --- due to low spin-splitting --- integer
filling factors $\nu = nh/eB$. These are accompanied by less pronounced lines away from integer
$\nu$, which are no longer perfectly parallel to $\nu$. Their number increases
with increasing magnetic field, i.e.\ the number of strongly localised states
increase with $B$ as expected in the single-particle picture \cite{Pru87}.\\
In contrast, we find that in the HF-interacting system, the number of
incompressible lines parallel to integer $\nu$ remains independent of $B$.
They are exhibited only within a strip of constant density around integer filling factor. In this region
of a nearly full (or empty) Landau level, the excess electron density is restricted by
the band edges and becomes too low to provide a complete screening
of the impurity density. The electron density is torn apart and the formation of
charge or hole islands leads to the observed charging lines in the compressibility \cite{PerC05,Efr88}.\\
This is in good agreement with the experimental results of Ref.\ \cite{IlaMTS04}.
In addition, we observe spin-splitting enhancement resulting in incompressible
strips also along {\em odd} integer filling factors.
We also observe lines of high contrast at $\nu \sim 1/2$ and $3/2$. At present,
these are unexplained and might be a numerical artifact.\\ \\
In our conductivity results, we see maxima aligning with integer $\nu$ even for
the non-interacting system. This is reminiscent of the results of Ref.\
\cite{CobBF99}. However, the fidelity of extrema of $\sigma_{xy}$ appears to be
lower. Results for the interacting system are currently in progress and will be presented
in a following publication.\\ \\
In conclusion, we have presented results from numerical calculations on Hall
conductance and electronic compressibility as a function of both magnetic field
and electron density and compared these to recent imaging experiments
\cite{CobBF99,IlaMTS04}.

%%%%%%%%%%%%%%%%%%%%%%%%%%%%%%%%%%%%%%%%%%%%%%%%%%%%%%%%%%%%%%%%%%%%%%%%%%%%%%%
%%
%% ACKNOWLEDGEMENT
%%
%%%%%%%%%%%%%%%%%%%%%%%%%%%%%%%%%%%%%%%%%%%%%%%%%%%%%%%%%%%%%%%%%%%%%%%%%%%%%%%
\begin{acknowledgement}
  We thank A.\ Croy, N.\ d'Ambrumenil and B. Huckestein for useful discussions.
\end{acknowledgement}
%%%%%%%%%%%%%%%%%%%%%%%%%%%%%%%%%%%%%%%%%%%%%%%%%%%%%%%%%%%%%%%%%%%%%%%%%%%%%%%

%%%%%%%%%%%%%%%%%%%%%%%%%%%%%%%%%%%%%%%%%%%%%%%%%%%%%%%%%%%%%%%%%%%%%%%%%%%%%%%
%%
%% BIBLIOGRAPHY
%%
%%%%%%%%%%%%%%%%%%%%%%%%%%%%%%%%%%%%%%%%%%%%%%%%%%%%%%%%%%%%%%%%%%%%%%%%%%%%%%%
%\bibliographystyle{prsty}\bibliography{bibliography/bibliograph}

\begin{thebibliography}{10}

\bibitem{CobBF99}
D.~H. Cobden, C.~H.~W. Barnes, and C.~J.~B. Ford, Phys. Rev. Lett. {\bf 82},
  4695  (1999).

\bibitem{IlaMTS04}
S. Ilani {\it et~al.}, Nature {\bf 427},  328  (2004).

\bibitem{KliDP80}
K.~v. Klitzing, G. Dorda, and M. Pepper, Phys. Rev. Lett. {\bf 45},  494
  (1980).

\bibitem{Pru87}
A.~M.~M. Pruisken,  in {\em The Quantum Hall Effect}, edited by R.~E. Prange
  and S.~M. Girvin (Springer, Berlin, 1987).

\bibitem{HucK90}
B. Huckestein and B. Kramer, Phys. Rev. Lett. {\bf 64},  1437  (1990).

\bibitem{KocHKP91b}
S. Koch, R.~J. Haug, K. v.~Klitzing, and K. Ploog, Phys. Rev. Lett. {\bf 67},
  883  (1991).

\bibitem{StrK05}
A. Struck and B. Kramer,   (2005), {arXiv}: cond-mat/0502095.

\bibitem{PerC05}
A. Pereira and J. Chalker,   (2005), {arXiv}: cond-mat/0502304.

\bibitem{Aok79}
H. Aoki, J. Phys. C {\bf 12},  633  (1979).

\bibitem{MacG88}
A. MacDonald and S. Girvin, Phys. Rev. B {\bf 38},  6295  (1988).

\bibitem{MacA86}
A. MacDonald and G. Aers, Phys. Rev. B {\bf 34},  2906  (1986).

\bibitem{KubMH65}
R. Kubo, S. Miyake, and H. Hashitsume,  in {\em Solid State Physics}, edited by
  F. Seitz and D. Turnbull (Academic Press, New York, 1965), Vol.~17, p.\ 269.

\bibitem{YanMH95}
S.-R.~E. Yang, A.~H. MacDonald, and B. Huckestein, Phys. Rev. Lett. {\bf 74},
  3229  (1995).

\bibitem{Efr88}
A. Efros, Solid State Commun. {\bf 65},  1281  (1988).

\end{thebibliography}

\end{document}